\documentclass[sigconf]{acmart}

\AtBeginDocument{%
  \providecommand\BibTeX{{%
    \normalfont B\kern-0.5em{\scshape i\kern-0.25em b}\kern-0.8em\TeX}}}

\copyrightyear{2020}
\acmYear{2020}
\setcopyright{rightsretained}
\acmConference[MODELS '20 Companion]{ACM/IEEE 23rd International Conference on Model Driven Engineering Languages and Systems}{October 18--23, 2020}{Virtual Event, Canada}
\acmBooktitle{ACM/IEEE 23rd International Conference on Model Driven Engineering Languages and Systems (MODELS '20 Companion), October 18--23, 2020, Virtual Event, Canada}\acmDOI{10.1145/3417990.3420057}
\acmISBN{978-1-4503-8135-2/20/10}

\begin{document}

\title{From Things' Modeling Language (ThingML) to Things' Machine Learning (ThingML2)}

\author{Armin Moin}
\affiliation{%
  \institution{Technical University of Munich}
 \country{Germany}
}
\email{moin@in.tum.de}

\author{Stephan R{\"o}ssler}
\affiliation{%
  \institution{Software AG}
  \city{Munich}
  \country{Germany}}
\email{stephan.roessler@softwareag.com}

\author{Marouane Sayih}
\authornote{Alumnus}
\affiliation{%
  \institution{Technical University of Munich}
  \country{Germany}
}
\email{marouane.sayih@mytum.de}

\author{Stephan G{\"u}nnemann}
\affiliation{%
 \institution{Technical University of Munich}
 \country{Germany}}
\email{guennemann@in.tum.de}

\renewcommand{\shortauthors}{Moin et al.}

\begin{abstract}
In this paper, we illustrate how to enhance an existing state-of-the-art modeling language and tool for the Internet of Things (IoT), called ThingML, to support machine learning on the modeling level. To this aim, we extend the Domain-Specific Language (DSL) of ThingML, as well as its code generation framework. Our DSL allows one to define things, which are in charge of carrying out data analytics. Further, our code generators can automatically produce the complete implementation in Java and Python. The generated Python code is responsible for data analytics and employs APIs of machine learning libraries, such as Keras, Tensorflow and Scikit Learn. Our prototype is available as open source software on Github.
\end{abstract}

\ccsdesc[500]{Software and its engineering~Domain specific languages}
\ccsdesc[300]{Software and its engineering~Software development techniques}
\ccsdesc[500]{Computing methodologies~Machine learning}

\keywords{domain-specific modeling, machine learning, internet of things}


\maketitle

\section{Introduction}
Existing modeling tools are not adequate for the era of the Internet of Things (IoT) and big data. First, the state-of-the-art tools for model-driven development of IoT services, such as ThingML \cite{ThingML, Harrand+2016} do not address Data Analytics (DA) needs. Second, workflow designers and frameworks for DA, such as KNIME come short of supporting the software development process beyond DA practices. Thus, we need a holistic approach to bridge the gap between Computer-Aided Software Engineering (CASE) on the one hand, and Computer-Aided Data Engineering and Analytics (CADEA) on the other hand. In this paper, we focus on Model-Driven Software Engineering (MDSE) from the former (CASE), and Machine Learning (ML) from the latter (CADEA). Moreover, we are particularly interested in the vertical application domains, which are related to the IoT and smart Cyber-Physical Systems (CPS).

\section{Background}\label{background}
\paragraph{Models} In Software Engineering (SE), \textit{models} often refer to the abstract artifacts concerning certain aspects of software systems, e.g., architecture or behavior. We call those artifacts \textit{software models} or \textit{model instances}. However, in DA, models are usually artifacts, which somehow abstract the observed data instances. For instance, according to Leskovec et al. \cite{Leskovec+2014}, one may define a model for a dataset (or data stream) as the underlying probability distribution, from which the observed data is presumably drawn. This is called the statistical approach. Alternatively, a model can be a summarization, an approximation or feature-based (i.e., representing the most extreme examples in data). We call such models \textit{data models}, \textit{DA models} or \textit{ML models}.

\paragraph{Model-based} One way of categorizing ML approaches is to group them into \textit{model-based} and \textit{instance-based} (also known as \textit{memory-based}) \cite{Bishop2006}. The difference between those two categories is that the so-called model-based approaches, e.g., Artificial Neural Networks (ANN), do not require access to the training data instances, once the ML model is trained. In contrast, instance-based approaches, such as Support Vector Machines (SVM) need to keep at least a subset of the training data even after training is accomplished. In this work, we do not consider this notion of model-based (vs. instance-based). Hence, our use of the term \textit{model-based} is concerned with the SE paradigm MDSE.

\section{Related work}
 \paragraph{Infer.NET} In 2013, Bishop \cite{Bishop2013} introduced the concept of \lq{}model-based ML\rq{} and proposed the framework \textit{Infer.NET} for probabilistic programming. The user should use the DSL of Infer.NET to define his or her data models, in this case, Probabilistic Graphical Models (PGM). Then, the framework would generate the software solution for the desired target application out of the PGM. Thus, for the first time, the term \textit{model-based} was used in the ML literature in the sense that we use it in the SE community. Prior to that, model-based had a different meaning in ML, as stated in Section \ref{background}.

 \paragraph{ThingML} ThingML \cite{ThingML, Harrand+2016} is a DSL and modeling framework for creating Heterogeneous and Distributed (HD) Services for the IoT and smart CPS applications. \textit{Things} in ThingML communicate with each other using asynchronous message passing. The user of ThingML shall define the following items: (i) The ports for message passing, the messages w/o parameters, and the properties (i.e., variables) for each \textit{thing}. (ii) The UML-like state machine (statechart), which models the behavior of each \textit{thing}. For this purpose, ThingML offers an imperative action language to support event-driven programing on the modeling level. (iii) The configuration for the entire application / service, which comprises the instances of the defined things and the connections among their specified ports.

\section{Proposed Approach: ThingML2}
\begin{figure}[h!]
	\includegraphics[width=0.5\textwidth]{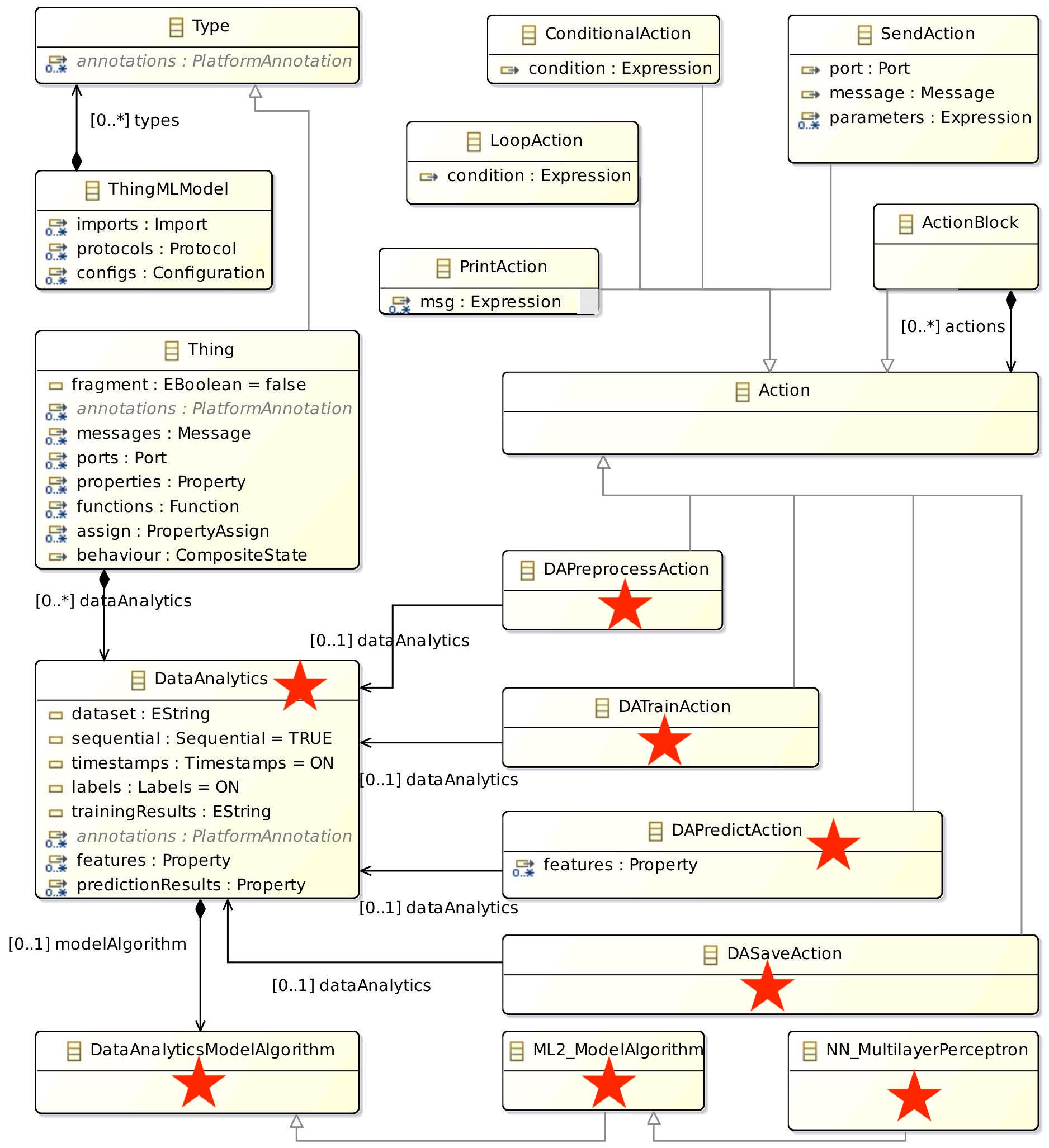}
	\caption{Part of the abstract syntax of the modeling language (meta-model). The red stars mark some of our extensions to the meta-model of ThingML.}
	\label{fig:ML2_Class_Diagram_Part}
\end{figure}

\begin{figure}[h!]
	\includegraphics[width=0.4\textwidth]{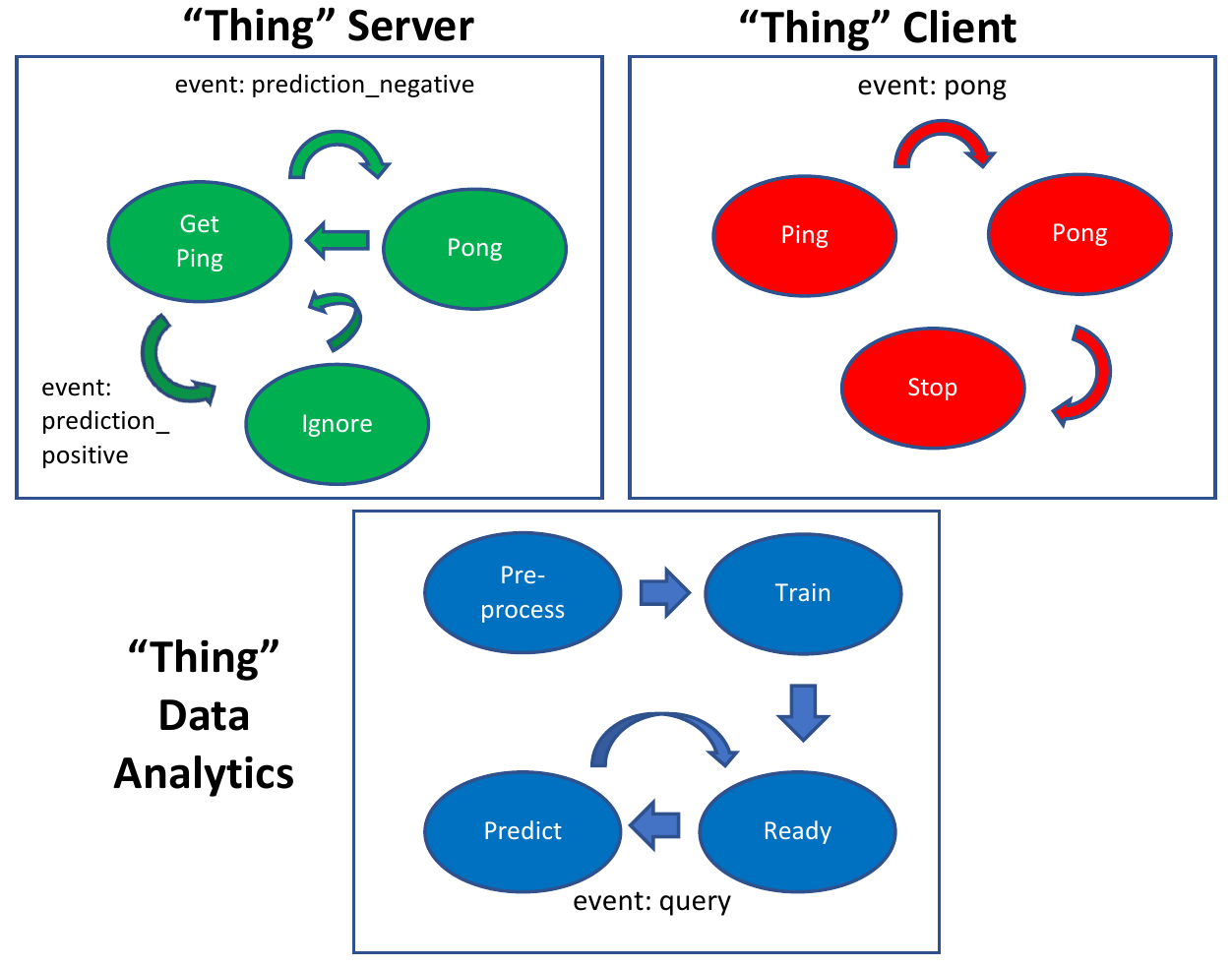}
	\caption{A sample use case: Smart Ping-Pong}
	\label{fig:SmartPingPong_AllInOne}
\end{figure}

In contrast to Infer.NET, where DA models were simultaneously used as software models too, we enhance software models to become capable of producing DA models and/or interacting with them. To this aim, we augment ThingML with ML functionalities on the modeling level. Concretely, ThingML users can employ our DSL to connect their software model instances to a range of ML models and algorithms via the APIs of the open source Python libraries and frameworks for ML, namely Keras, TensorFlow and Scikit Learn. Similar to the methodology of ThingML, we also support full generation of the source code (in our case also Python) out of the software model instances in an entirely automated manner. Figure \ref{fig:ML2_Class_Diagram_Part} shows part of the architecture of the abstract syntax of our modeling language. In addition, we demonstrate a sample use case using state machines in Figure \ref{fig:SmartPingPong_AllInOne}, where an existing example of the ThingML project for client-server interaction, called PingPong, is enhanced to prevent Distributed Denial of Service (DDOS) attacks. Thus, we refer to that as \textit{Smart PingPong}. Note that \textit{Thing Client} remained the same as the original example of ThingML. However, \textit{Thing Server}, which previously only had a single state, i.e., \textit{Active} (not shown there), has now three states as depicted in Figure \ref{fig:SmartPingPong_AllInOne}, among which it can switch, based on the predictions of the newly introduced thing, \textit{Data Analytics}. For every \textit{thing}, which shall perform DA tasks, the user must specify a DA block in the model instance, in addition to the above-mentioned items, e.g., ports, messages and state machine. The DA block was not offered in ThingML. We call our proposed approach ThingML2. Our prototype, called ML-Quadrat, is available at \url{https://github.com/arminmoin/ML-Quadrat}.

\begin{acks}
This extended abstract accompanies the authors' poster at MODELS 2020. This work is partially funded by the German Federal Ministry of Education and Research (BMBF) through the Software Campus initiative (project ML-Quadrat).
\end{acks}

\bibliographystyle{ACM-Reference-Format}
\bibliography{refs}


\end{document}